\documentclass[11pt]{article}
\usepackage{cospar}
\usepackage{url}

\usepackage{graphicx}
\usepackage[figuresright]{rotating}

\def\aj{{\em Astron. J.}}
\def\apj{{\em Astrophys. J.}}
\def\apjs{{\em Astrophys. J. Suppl.}}
\def\aea{{\em Astron. \& Astrophys.}}
\def\mnras{{\em Mon. Not. Roy. Ast. Soc.}}
\def\an{{\em Astron. Nach.}}
\def\pasj{{\em Pub. Ast. Soc. Jap.}}

\def\cgs{erg cm$^{-2}$ s$^{-1}$ }
\def\norm{keV s$^{-1}$ sr$^{-1}$ cm$^{-2}$ keV$^{-1}$ } 
\def\lum{erg s$^{-1}$ }
\def\nh{cm$^{-2}$ }

\title{THE X-RAY BACKGROUND AND THE DEEP X-RAY SURVEYS}

\author{R. Gilli\address{Istituto Nazionale di Astrofisica (INAF),
Osservatorio Astrofisico di Arcetri, Largo E. Fermi 5, 50125 Firenze,
Italy}}

\begin{document}

\maketitle

\begin{abstract}

The deep X-ray surveys performed by the two major X-ray observatories
on flight, Chandra and XMM, are being resolving the bulk of the cosmic
X-ray background (XRB) in the 2--10 keV energy band, where the sky
flux is dominated by extragalactic emission. Although the actual
fraction depends on the absolute sky flux, which is measured with an
uncertainty of $\sim 40\%$, most of the XRB is already
resolved. Optical identifications of the X-ray sources in the deep
surveys are being showing that these are mainly AGN, most of which
being obscured as predicted by population synthesis models. However,
first results indicate that the redshift distribution of the sources
making the XRB seems to peak at much lower redshift than predicted by
the models. In this article I will briefly review and discuss the
measurements of the XRB spectrum and the AGN synthesis models of the
XRB. Then, I will introduce the Chandra and XMM deep X-ray surveys,
mainly focusing on the Chandra Deep Field North and South.  Finally,
the properties of the X-ray sources populating the deep surveys will
be described and compared with the predictions of the most recent
synthesis models.
\end{abstract}

\section*{INTRODUCTION}

The origin of the X-ray background in the 2-10 keV energy range has
been finally understood. The extremely deep surveys by Chandra in the
Deep Field South (Giacconi et al. 2002; Rosati et al. 2002) and Deep
Field North (Brandt et al. 2001; Barger et al. 2002) have shown that
most, if not all, the XRB emission in that energy range has been
resolved into single sources. In particular, the optical and X-ray
properties of the sources in the deep fields are showing that the main
contribution is provided by a population of obscured AGN. This
confirms the main prediction of population synthesis models
(e.g. Setti \& Woltjer 1989; Comastri et al. 1995; Fabian et al. 1999;
Gilli, Salvati \& Hasinger 2001), which explain the XRB spectrum in
the $\sim 1-100$ keV band as the emission integrated over cosmic time
of unobscured and obscured AGN, the latter being more numerous by a
factor of $\sim 4-10$. Once the origin of the 2-10 keV XRB is finally
explained, the results from the deep surveys, combined with synthesis
models, should put tighter constraints on the cosmological properties of
the sources making the XRB, like their X-ray luminosity function and
evolution (especially those of obscured AGN, which are still
completely unknown), as well as their average obscuration as a
function of redshift and luminosity.

\begin{figure}[t]
\begin{center}
\includegraphics[width=140mm]{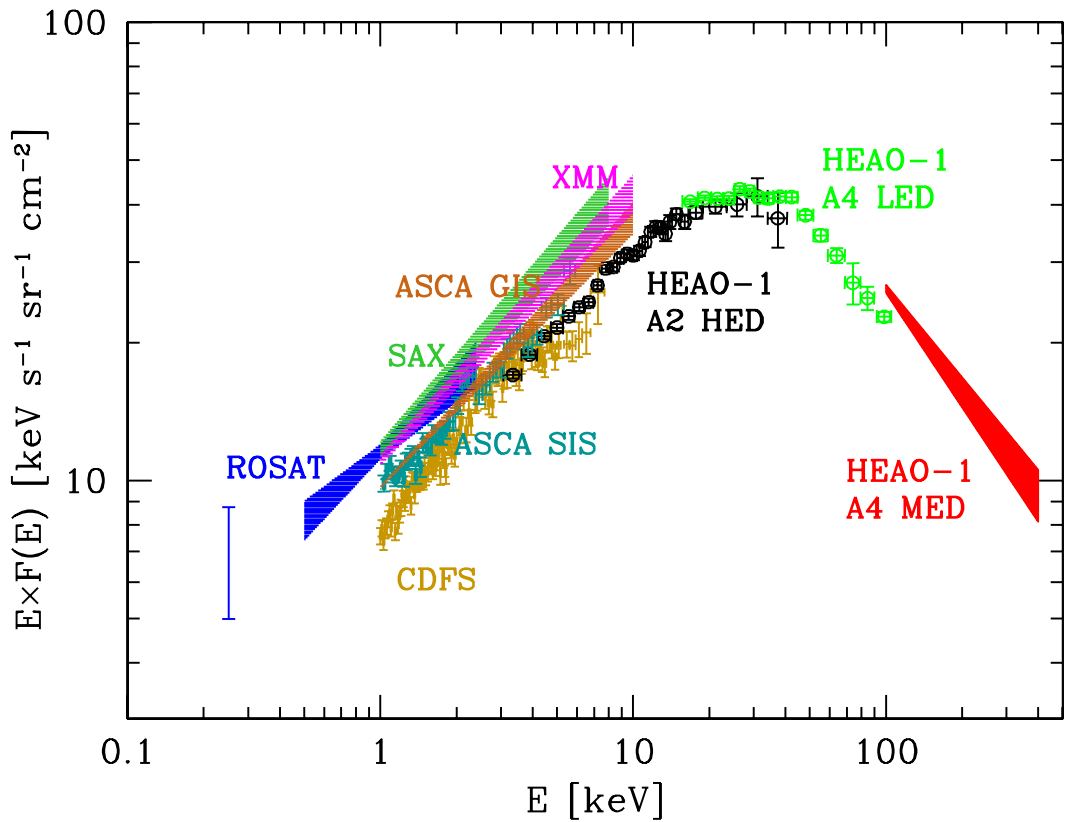}
\caption{The extragalactic X-ray background spectrum from 0.2 to 400
keV. Different colors correspond to measurements by different
missions/instruments as labeled. The reference list for the shown data
is the following: ROSAT 0.25 keV (Warwick \& Roberts 1998); ROSAT
0.5-2.4 keV (Georgantopoulos et al. 1996); HEAO-1 A2 HED + A4 LED
(Gruber 1992; Gruber et al. 1999); HEAO-1 A4 MED (Kinzer et al. 1997);
SAX (Vecchi et al. 1999); ASCA SIS (Gendreau et al. 1995); ASCA GIS
(Kushino et al. 2002); XMM (Lumb et al. 2002); CDFS (Tozzi et
al. 2001a).}
\end{center}
\end{figure}

\section*{THE COSMIC X-RAY BACKGROUND MEASUREMENTS}

At present, in the 10--100 keV range, where the bulk of the XRB energy
resides, the only available measurements are those performed in the
late 1970s by HEAO-1 (Marshall et al. 1980; Gruber 1992). The data
showed that the XRB spectrum has a characteristic ``bell'' shape
peaking at $\sim 30-40$ keV, which, at lower energies, can be
approximated by a power-law with photon index $\Gamma=1.4$. A recent
reanalysis of HEAO-1 data by Gruber et al. (1999) confirmed those
earlier results, but also showed that the calibration uncertainties
among HEAO-1 detectors are of the order of 10\% in the overlapping
energy bands (Fig.~1). The situation is more complicated at lower
energies where a number of measurements have been obtained. While the
spectral slope of the 2--10 keV XRB has always been found to have
small variations around $\Gamma=1.4$, there are significant
discrepancies in the spectral normalization. From the highest (11.5
\norm; SAX, Vecchi et al. 1999) to the lowest measured value (8 \norm;
HEAO-1, Marshall et al. 1980) the variation is of the order of
$\sim 40\%$. The most recent measuments have been obtained by Kushino et
al. (2002) from a combination of 91 ASCA GIS fields, with a total sky
coverage of 50 deg$^2$, and by Lumb et al. (2002) from a combination
of 8 XMM pointings covering 6.5 deg$^2$ in total. The XRB spectral
slope is found to be $\Gamma=1.412\pm 0.032$ and $\Gamma=1.42 \pm
0.03$ in the ASCA and XMM data, respectively (errors are at $90\%$
confidence level), again very close to $\Gamma=1.4$. The power-law
normalizations ($9.66 \pm 0.07$ and $11.1\pm 0.32$ \norm for the ASCA
and XMM data, respectively, including the resolved bright sources) are
within the range already spanned by previous missions, but do not
suggest that the measurements of the XRB intensity are converging to a
well constrained value. Interestingly, all the measurements of the
2-10 keV XRB intensity since HEAO-1 are higher than the HEAO-1 value.

\section{AGN POPULATION SYNTHESIS MODELS}

In 1989, when the resolved XRB fraction was just a few percent, Setti
\& Woltjer (1989) proposed that the flat slope of the 2--10 keV XRB
was due to a population of obscured AGN in addition to the bright
unobscured AGN with steep spectra observed at that time. Since then, a
number of models have been worked out and refined constantly (Madau,
Ghisellini \& Fabian 1994; Comastri et al. 1995; Pompilio, La Franca
\& Matt 1999; Wilman \& Fabian 1999). With the increasing sensitivity
of recent X-ray surveys, the population of obscured AGN emerged (see
e.g. Mushotzky et al. 2000), and the main prediction of the AGN
synthesis models was then confirmed. However, given the large number
of uncertain parameters involved, it is crucial to check synthesis
models against the largest number of observational constraints. The
main uncertainties are related to unobscured AGN. While the X-ray
luminosity function (XLF) and evolution of unobscured AGN are rather
well known (e.g. Miyaji, Hasinger \& Schmidt 2000), nothing is known
about obscured AGN. Comastri et al. (1995) showed that, assuming for
the obscured AGN a distribution of absorbing columns and the same
evolution and XLF (upscaled by a factor of a few) of unobscured ones,
it was possible to fit the full XRB spectrum in the $\sim 1 - 100$ keV
band, as well as the integral counts (logN-logS relation) in the 0.5-2
keV and 2-10 keV energy bands. Also, the redshift distribution of AGN
detected in ROSAT and HEAO-1 samples (at limiting fluxes of
$f_{0.5-2}\sim 10^{-14}$ \cgs and $f_{2-10}\sim 3\:10^{-11}$ \cgs,
respectively) were in good agreement with the model
expectations. Later, more constraints became available. The local
ratio between obscured and unobscured AGN (in the Seyfert luminosity
regime) was found to be $\sim 4$ (Maiolino \& Rieke 1995), and the
column density distribution of local Seyfert 2s was determined
(Risaliti et al. 1999). Using these new constraints it was shown
(Gilli, Risaliti \& Salvati 1999) that, if the column density distribution in
obscured AGN is the same at all redshifts and luminosities, then
additional obscured sources at moderate/high redshifts are required to
match the XRB constraints. Gilli, Salvati \& Hasinger (2001, hereafter
GSH01) proposed that obscured sources evolve slightly faster than
unobscured ones. In particular they favored a model (model B) where
the ratio between obscured and unobscured AGN increases from 4 at
$z=0$ to 10 at $z=1.3$, where both populations stop evolving. This
model was able to reproduce the broad set of observational constraints
used by Comastri et al., furthermore extending the agreement to lower
fluxes. In particular, the redshift distribution of soft X-ray
selected AGN in the ROSAT Ultra Deep Survey (UDS, Lehmann et al. 2001)
at a limiting flux of $f_{0.5-2}\sim 10^{-15}$ \cgs was nicely
reproduced, as well as that of hard X-ray selected AGN in the ASCA
Large Sky Survey (Akiyama et al. 2000) at a limiting flux of
$f_{2-10}\sim 10^{-13}$ \cgs. Moreover, AGN synthesis models showed
that, while the XRB spectrum can be simply fitted with a population of
Seyfert 2 galaxies in addition to unobscured AGN, the ASCA and
BeppoSAX source counts in the hard band are matched only by assuming a
population of luminous obscured AGN, the so-called QSO2s (GSH01;
Comastri et al. 2001). It is here noted that the main contribution to
the source counts is expected to be produced by QSO2s with column
densities in the range log$N_H=22-23$, while the contribution of
Compton thick QSO2s is expected to be negligible.


\begin{figure}
\begin{minipage}{90mm}
\includegraphics[width=90mm]{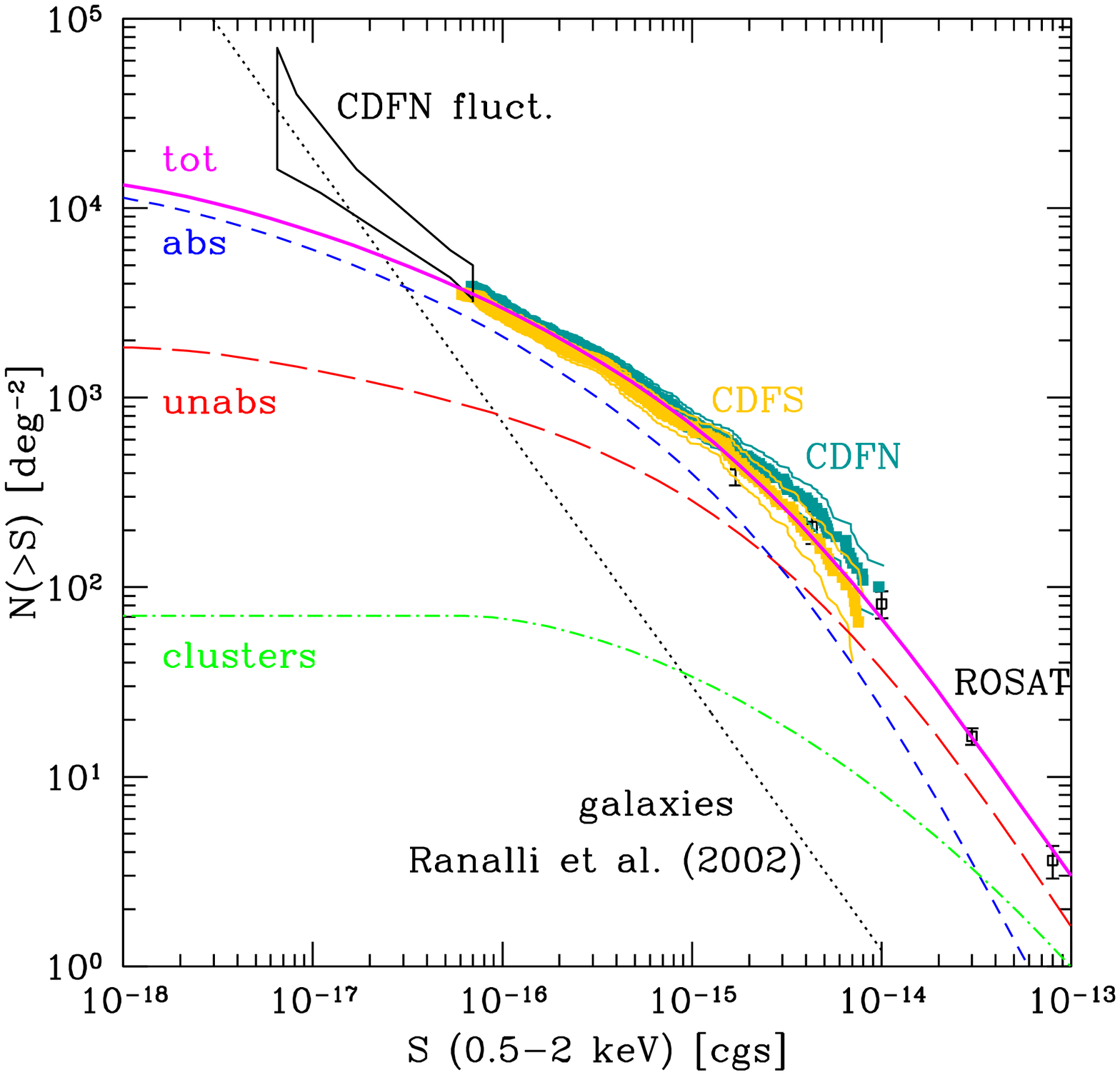}
\end{minipage}
\hfil\hspace{\fill}
\begin{minipage}{90mm}
\includegraphics[width=90mm]{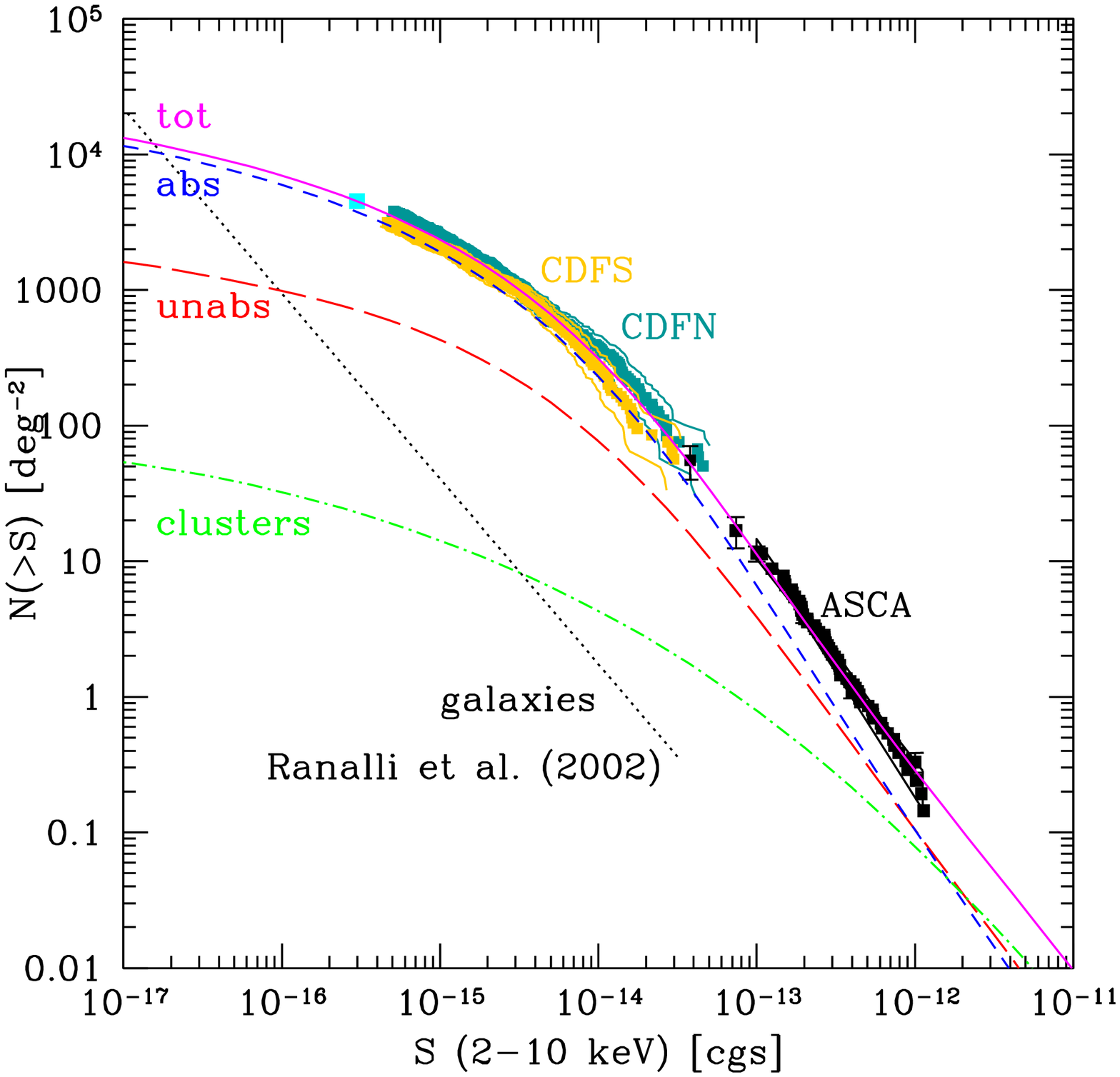}
\end{minipage}
\caption{Soft ($left$) and hard ($right$) logN-logS compared with the
predictions of model B by GSH01. Different curves correspond to the
contribution of different classes of objects as labeled. The galaxy
counts predicted by the Ranalli et al. (2002) model are also plotted
as a dotted line. CDFS and CDFN data are from Rosati et al. (2002) and
from Brandt et al. (2001). The CDFN fluctuation analysis box is
adapted from Miyaji \& Griffiths (2002). ROSAT data are from Miyaji et
al. (2000). ASCA data (black squares) at $f_{2-10}=4\: 10^{-14}$, $7\:
10^{-14}$ and $>10^{-13}$ \cgs are from Ogasaka et al. (1998), Ueda
(2001) and Cagnoni et al. (1998), respectively. The deepest datapoint
in the hard logN-logS is from Moretti et al. (2002).}
\end{figure}

\begin{table}
\begin{tabular}{lcclccrc}
\hline\hline
Survey& RA& DEC& X-ray data& $f_{0.5-2}^a$& $f_{2-10}^b$& $N^c$& Reference\\
& \multicolumn{2}{c}{(J2000)}& & & & & \\
\hline
CDFN& 189.200&  62.231&	2 Ms Chandra ACIS-I& 0.1&  0.1&	503&  1\\
    &&& 			 180 ks XMM& 4&    2&	$\sim 200$&  2\\		
CDFS& 53.116&  -27.808&	1 Ms Chandra ACIS-I& 0.5&  0.4&	346& 3,4\\
    & 	&&		 370 ks XMM&    &   &	   &  5\\
Lockman Hole& 163.179&	57.480&	300 ks Chandra HRC&    &   &	   &  \\ 	  
&	&& 			100 ks XMM& 3.1& 1.4&	$\sim 200$&  6\\	 
Lynx Field&	132.229& 44.909&     180 ks Chandra ACIS-I& 1.7& 1.3&	153&  7\\ 
&	&&		 140 ks XMM&	  &   &	   &  \\
Groth Strip& 214.429& 52.474& 200 ks Chandra ACIS-I&  &   &	   &  \\ 
&		&&	  80 ks XMM&   6& 4&	$\sim 150$ & 2 \\
13hr field&	203.654& 37.912& 120 ks Chandra ACIS-I& & & 214& 8\\
	&	&&		 130 ks XMM&	5& $\sim 3$& 216& 8 \cr
\hline
\end{tabular}
\vskip -0.8cm
\caption{Summary of the main deep and moderately deep Chandra/XMM surveys.
\newline $^a$ Limiting flux in the 0.5-2 keV band in units of $10^{-16}$ \cgs.
\newline $^b$ Limiting flux in the 2-10 keV band in units of $10^{-15}$ \cgs.
\newline $^c$ Number of detected sources.
\newline References: 1)Alexander at al. (2002); 2)Miyaji et al. (2002);
3)Giacconi et al. (2002); 4)Rosati et al. (2002); 5)Hasinger et
al. (2002); 6)Hasinger et al. (2001); 7)Stern et al. (2002); 8)Page et
al. (2002).}
\end{table}

\section{THE DEEP X-RAY SURVEYS}

A number of deep ($\sim$ Ms) and moderately deep ($\sim 200$ ks)
X-ray surveys are being conducted in different sky fields with Chandra
and XMM.

The Chandra Deep Field North (CDFN, Brandt et al. 2001) and the
Chandra Deep Field South (CDFS, Giacconi et al. 2002; Rosati et
al. 2002) have been observed with the ACIS-I array for 2 Ms and 1 Ms
respectively, and represent the two deepest X-ray surveys to date. In
the 2 Ms CDFN the achieved sensitivity limit is $\sim 1.5\: 10^{-17}$
\cgs in the soft band and $\sim 1.0\: 10^{-16}$ \cgs in the hard band
(Alexander et al. 2002). In the CDFS the sensitivity limits of
$f_{0.5-2}\sim 5.5\: 10^{-17}$ \cgs and $f_{2-10}\sim 4.5\: 10^{-16}$
\cgs obtained by Rosati et al. (2002) have been pushed to
$f_{0.5-2}\sim 2.4\: 10^{-17}$ \cgs and $f_{2-10}\sim 2.1\: 10^{-16}$
\cgs by refining the detection techniques (Moretti et al. 2002). The
Chandra ACIS-I observations of these fields have been complemented
with deep XMM exposures: 180 ks and 370 ks of clean XMM data have been
obtained for the CDFN (Miyaji et al. 2002) and the CDFS (Hasinger et
al. 2002), respectively. The full source catalogs of the 1 Ms ACIS-I
observations of the CDFN and CDFS have been already released (Brandt
et al. 2001; Giacconi et al. 2002), while the additional 1 Ms ACIS-I
observation of the CDFN and the XMM observations of the CDFN and CDFS
are actually under analysis and only preliminary results are available
(Alexander et al. 2002; Miyaji et al. 2002; Hasinger et
al. 2002). Deep Chandra HRC and XMM exposures are on going in the
Lockman Hole (see Hasinger et al. 2001 for the analysis of the first
XMM data), where the deepest ROSAT and ASCA surveys were already
performed (the ROSAT UDS, Lehmann et al. 2001, and the ASCA Deep
Survey, Ishisaki et al. 2001). Other examples of combined Chandra/XMM
deep surveys are being carried out in the Lynx field (Stern et
al. 2002), in the Groth-Westphal strip (Miyaji et al. 2002) and in the
13hr field (Page et al. 2002). A summary of the main deep and
moderately deep Chandra/XMM surveys is shown in Table~1.

In the soft band, the deepest observations are detecting sources about
two orders of magnitude fainter than those observed by the ROSAT
UDS. The progress in the hard X-rays is even higher, since Chandra is
now detecting sources $\sim 300$ times fainter than those detected in
the ASCA Deep Survey.

As shown by Tozzi et al. (2001a), the integrated emission of all the
CDFS sources has already resolved the 2-10 keV background flux
measured by HEAO-1. Furthermore, the stacked spectrum of the whole
CDFS sample is well described by a powerlaw with $\Gamma=1.4$ (Tozzi
et al.  2001a), in excellent agreement with the measurements of the XRB
spectral slope in that energy range (see Fig.~1).

The soft and hard logN-logS calculated in the CDFS and CDFN are
confirming the predictions by AGN synthesis models. As shown in Fig~2,
the logN-logS curve for model B of GSH01, which is the summed
contribution of absorbed AGN, unabsorbed AGN and clusters of galaxies,
can reproduce the CDFN and CDFS data at all fluxes. Some deviations
are observed in the soft logN-logS when considering the data from the
CDFN fluctuation analysis (Miyaji and Griffiths 2002), which are
higher than the model expectations. However, as shown in Fig~2 (left),
at those faint fluxes a population of normal/starburst galaxies is
expected to provide a major contribution (Ptak et al. 2001; Ranalli,
Comastri \& Setti 2002).

\begin{figure}[t]
\begin{minipage}{90mm}
\includegraphics[width=90mm]{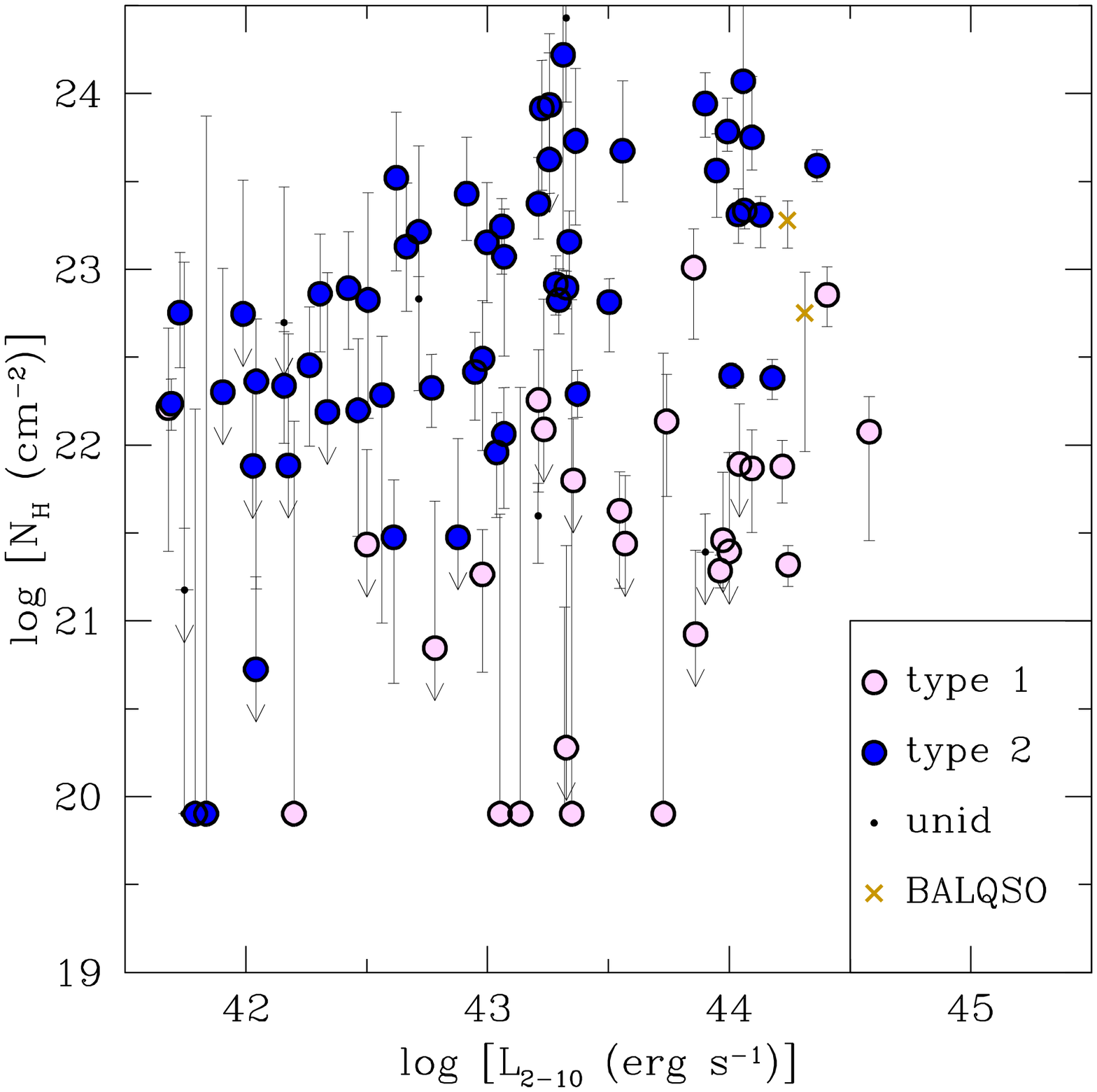}
\end{minipage}
\hfil\hspace{\fill}
\begin{minipage}{90mm}
\includegraphics[width=90mm]{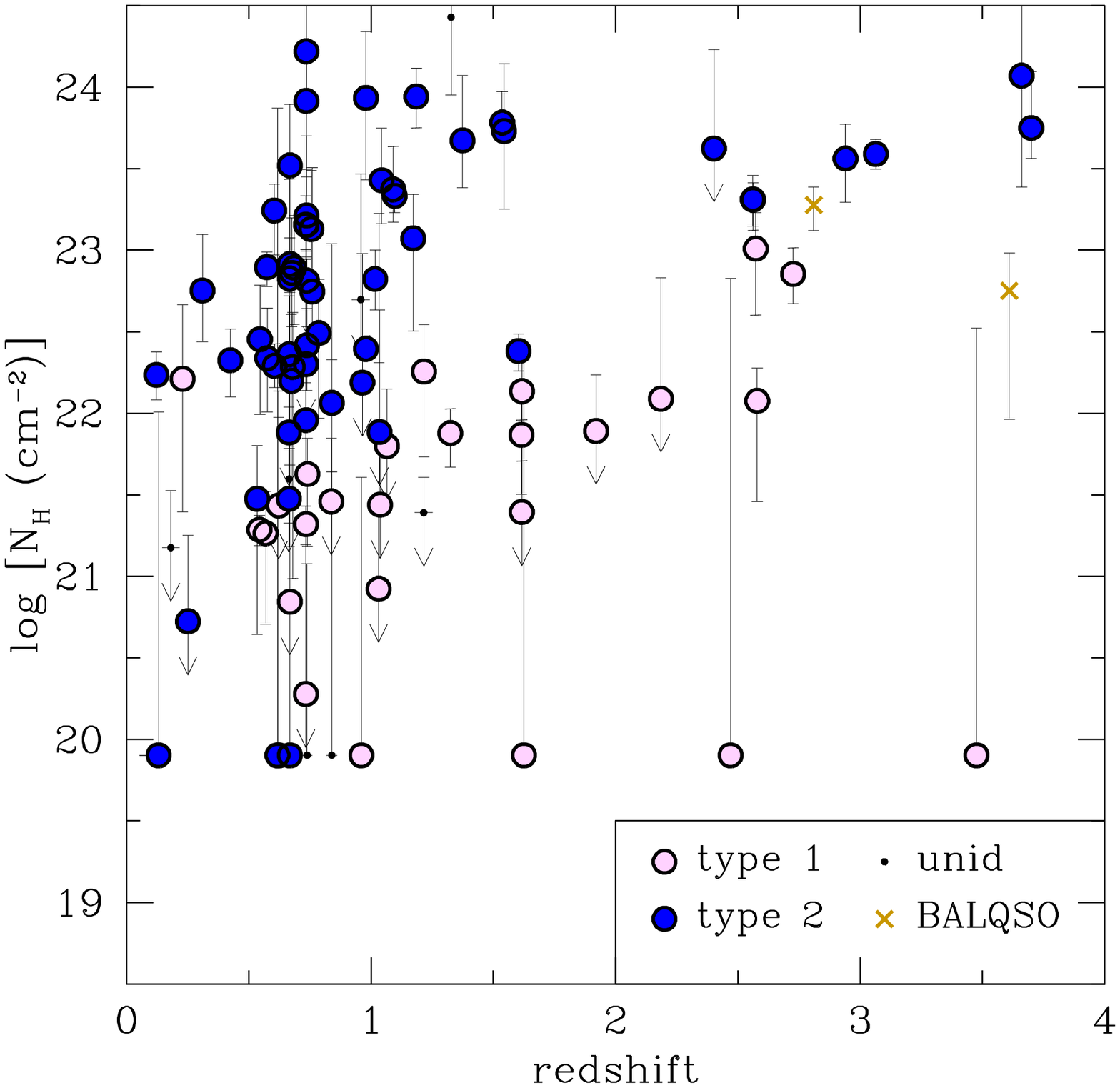}
\end{minipage}
\caption{$Left$: rest-frame absorbing column density versus rest
frame, intrinsic (de-absorbed) luminosity in the 2-10 keV band for the
CDFS sources with $f_{2-10}> 5 \: 10^{-16}$ \cgs (a cosmological model
with $H_0=50$ km s$^{-1}$ Mpc$^{-1}$ and $q_0=0.5$ has been
assumed). Different symbols represent different optical types as
labeled. Type 1 AGN do show broad lines, while type 2 AGN do not (see
Szokoly et al. in preparation for the detailed classification
criteria). Sources with uncertain classification are plotted as dark
dots. Obscured QSOs populate the upper-right corner of the plot (with
$L_{2-10}>10^{44}$\lum and $N_H>10^{22}$\nh). As already observed in
shallower surveys (e.g. HELLAS, Fiore et al. 2001) several broad line
AGN are found to have X-ray absorption in excess of
$10^{22}$\nh. $Right$: rest-frame absorbing column density versus
redshift for the same source sample. Symbols are the same as in the
left panel.}
\end{figure}

\subsection*{The sources populating the Deep Surveys}

Due to the extremely faint X-ray fluxes sampled by the deep surveys,
for standard X-ray to optical flux ratios, 8-10 meter class telescopes
are needed for the optical followup (see Hornschemeier 2002 for a
review). Optical spectra of the X-ray sources in the CDFN are being
obtained at Keck (Barger et al. 2002), while the VLT is primarily
exploited for optical spectroscopy of CDFS sources (Szokoly et al. in
preparation; see also Hasinger et al. 2002). The spectroscopic
completeness in the CDFS and CDFN is roughly $40-50\%$, but it
increases up to $\sim 80\%$ when considering the inner part of the
fields and sources with X-ray fluxes well above the survey sensitivity
limit. A direct redshift estimate is not possible for a significant
fraction of X-ray sources even with 10m class telescopes: $\sim 25\%$
of the X-ray sources have counterparts fainter than R=25, while $\sim
15\%$ do not have any counterpart down to R magnitudes of 26.1-26.7
(Giacconi et al. 2002). Nevertheless, photometric redshifts are being
determined for optically faint sources. The counterparts of
the X-ray sources reveal a broad variety of optical properties. Broad
line and narrow line AGN are commonly found, but there are also
sources with no obvious high-excitation emission lines in their
spectra. When the optical classification is uncertain, the X-ray
properties such as the X-ray luminosity and hardness ratio, in
addition to the X-ray to optical flux ratio ($f_x/f_{opt}$), can be used
to discriminate between nuclear and stellar activity (see
e.g. Hasinger et al. 2002). As shown by Hornschemeier et al. (2002)
and Giacconi et al. (2002), about $15-20\%$ of the sources populating
the deep X-ray surveys have $f_x/f_{opt}$ values typical of
normal/starbursts galaxies, with this fraction being higher in the
soft band and towards faint fluxes. On the contrary, the contribution
of non-active galaxies is negligible when selecting sources with 2-10
keV fluxes above a few $\:10^{-15}$ \cgs.

In Fig.~3 (left panel) the absorbing column of CDFS sources, measured
from the X-ray spectral fit, is plotted against their rest frame,
intrinsic (i.e. de-absorbed) X-ray luminosity in the 2-10 keV band
($L_{2-10}$), calculated assuming $H_0=50$ km s$^{-1}$ Mpc$^{-1}$ and
$q_0=0.5$.  When the photon statistics is low, the spectral index is
fixed to $\Gamma=1.8$ and the absorbing column measured
accordingly. Different symbols correspond to different optical types
(only sources with $f_{2-10}> 5 \: 10^{-16}$ \cgs and good
spectroscopic redshift are shown). A fraction of AGN with broad
optical lines are found to have X-ray absorption in excess of
$10^{22}$\nh, confirming the optical to X-ray classification mismatch
already observed in shallower X-ray surveys (e.g. HELLAS, Fiore et
al. 2001; Comastri et al. 2001). A number of sources are found to have
$L_{2-10}>10^{44}$\lum and $N_H>10^{22}$\nh, which can be well
considered the faint tail of the QSO2 population (a similar result has
been found in the Lockman Hole by Mainieri et al. 2002a). Given the
relatively low space density of luminous sources, the high-luminosity
tail of the QSO2 population is best sampled by shallow, wide area
surveys.

\begin{figure}[t]
\begin{minipage}{90mm}
\includegraphics[width=90mm]{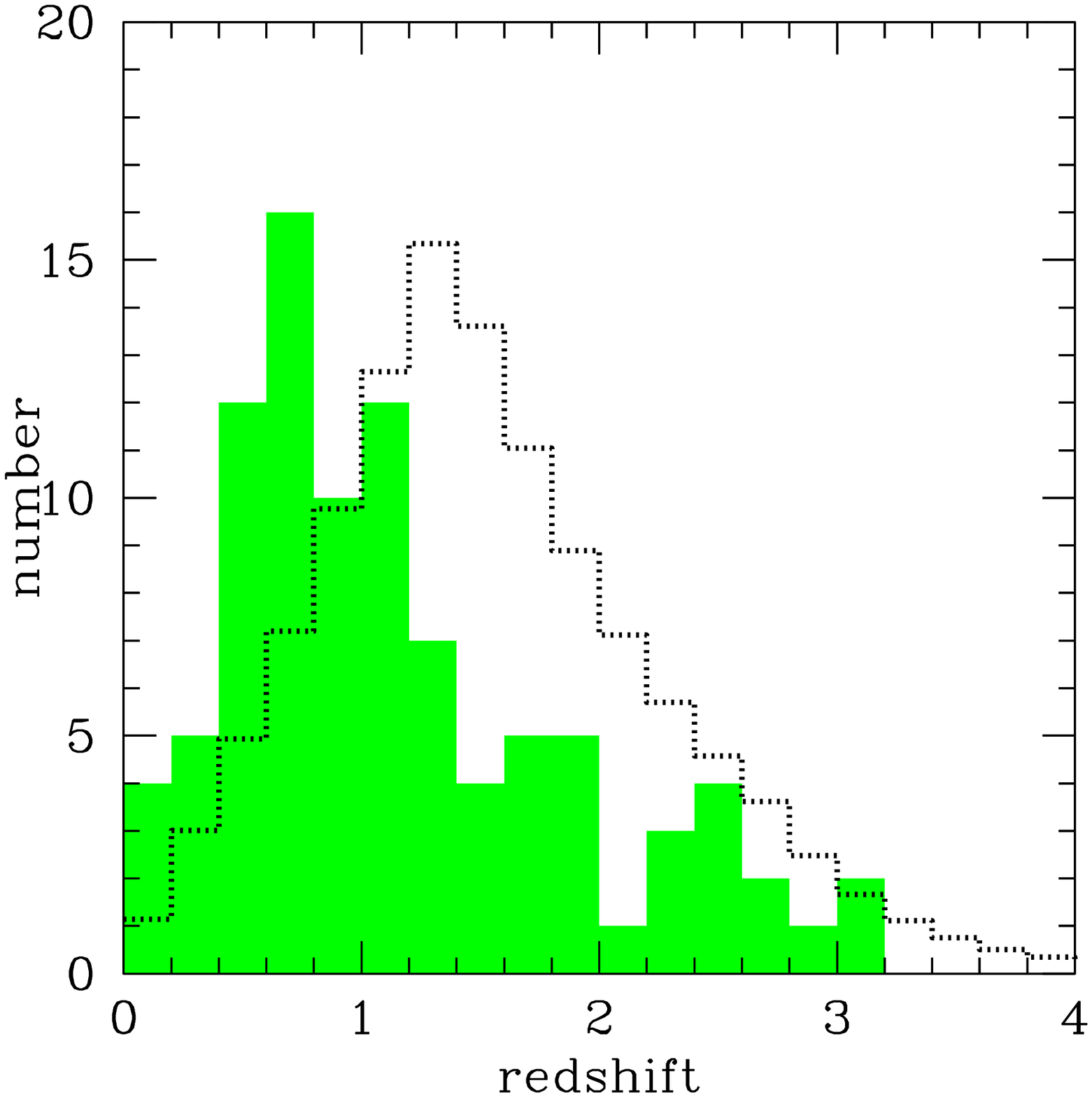}
\end{minipage}
\hfil\hspace{\fill}
\begin{minipage}{90mm}
\includegraphics[width=90mm]{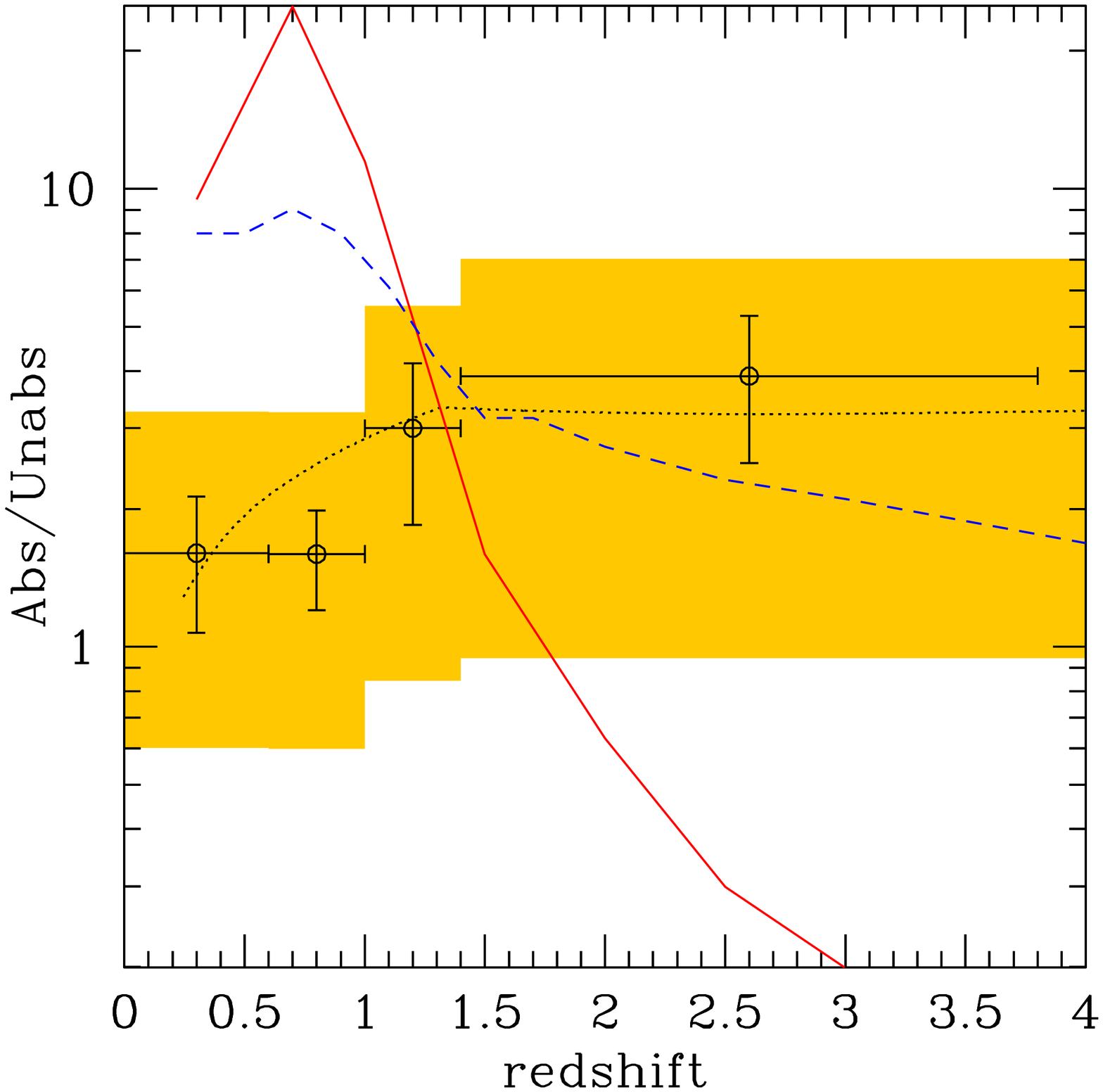}
\end{minipage}
\caption{$Left$: Redshift distribution for the 93 X-ray sources with
$f_{2-10}>5 \: 10^{-15}$ \cgs in the central regions of the CDFS,
CDFN, Lockman Hole, Lynx field, and SSA13 field (see text; only
spectroscopic redshifts are considered). With this selection the
achieved spectroscopic completeness is $\sim 80\%$. Sources belonging
to large scale structures in the CDFS and CDFN are excluded. The data
(shaded area) are compared with the predictions of model B by GSH01 at
the same limiting flux (dotted line), normalized to 116 sources
(93/116 = 80\%). A clear excess of sources is observed at $z<1$ with
respect to the model predictions.\newline $Right$: Ratio between AGN
with log$N_H>22$ and log$N_H<22$ as a function of redshift for the
sources with $f_{2-10}>5\: 10^{-16}$ \cgs detected in the inner
regions of the CDFS and CDFN (the spectroscopic completeness of this
sample is $\sim 60\%$). Redshift bins have been chosen to contain
approximately the same number of sources. The predictions for the
Franceschini et al. (2002) model and the Gandhi \& Fabian (2002) model
at comparable fluxes ($f_{2-10}=8\: 10^{-16}$ and $5\: 10^{-16}$ \cgs,
respectively) are also plotted as a solid and dashed line,
respectively. The shaded area shows the possible range covered by the
data under the two extreme assumptions that the unidentified sources
(40\% of the sample) are either all obscured or all unobscured (see
text). Both models seem to predict too many obscured AGN at $z<1$. For
comparison the prediction by model B of GSH01 at $f_{2-10}>5\:
10^{-16}$ \cgs is plotted as a dotted line.}
\end{figure}

\subsection*{The source redshift distribution}

Most of the sources identified in the CDFN, CDFS and Lockman Hole are
found to be at $z<1$ (see Fig.~10 of Barger et al. 2002 for the CDFN;
Fig.~9 of Tozzi et al. 2001b for the CDFS; Fig.~1 of Mainieri et
al. 2002b for the Lockman Hole). Synthesis models of the X-ray
background predict that at these faint fluxes the X-ray source
redshift distribution is peaked at $z \sim 1.3-1.5$, in contrast with
the observed data. A direct comparison between the published
distributions and the model predictions is however complicated by the
following effects:

1) optical identifications are still far from complete. In general,
   the still unidentified sources are expected to be at higher
   redshift than those already observed.

2) $\sim 15-20\%$ of the sources populating the deep surveys are
   non-active galaxies at $z<1$.

3) given the narrow field of view of the deep surveys (0.1-0.2
   deg$^2$) the redshift distribution could be dominated by the
   presence of large scale structures. This is indeed the case in the
   CDFS (Gilli et al. 2003), where narrow spikes in the source
   redshift distribution have been found at $z=0.67$ and $z=0.73$, and
   also in the CDFN, where redshift structures have been found at
   $z=0.84$ and $z=1.02$ (Barger et al. 2002).

Nevertheless, the spectroscopic completeness can be significantly
increased by selecting sources with an X-ray flux well above the
survey limit in those subregions with higher optical coverage. Also, if
only bright and hard X-ray selected sources are considered, the
percentage of non-AGN objects drops dramatically.
 
In Fig.~4 (left) it is shown the redshift distribution of the 93 sources with
$f_{2-10}>5\:10^{-15}$ \cgs in the central regions of the CDFS, CDFN,
Lockman Hole (Mainieri et al. 2002a), Lynx field (Stern et al. 2002)
and SSA13 field (Mushotzky et al. 2000). The spectroscopic
completeness of this combined sample is $\sim 80\%$. Sources belonging
to the large scale structures observed in the CDFS and CDFN have been
excluded. The redshift distribution predicted by model B of GSH01 at
the same limiting flux and normalized to 116 objects (93/116 = 80\%)
is also shown in Fig.~4 (left). An excess of sources at $z<1$ with respect to
the model predictions is still found in this clean sample.

\section{DO OBSCURED AGN EVOLVE LIKE STARBURST GALAXIES?}

A solution to this discrepancy has been recently proposed by
Franceschini, Braito \& Fadda (2002), who suggested that obscured AGN
undergo a very fast evolution up to $z=0.8$. The physical scenario
supporting this idea is that obscured AGN are related to the fast
evolving starburst population necessary to reproduce the ISO
mid-infrared counts (Franceschini et al. 2001). By shifting most of
the obscured AGN at $z<1$, that model nicely reproduces the redshift
distribution observed in the deep surveys; unobscured sources are
however still needed to explain the high redshift tail of the
distribution (see Fig.~5 in Franceschini et al. 2002). A more refined
model has been recently worked out by Gandhi \& Fabian (2002), who
also make a connection between obscured AGN with the infrared
starburst population. Even this model is able to reproduce the low-z
peak in the redshift distribution, with the main contribution at $z<1$
provided by obscured AGN. Both these new models are bound to predict a
decrease with redshift in the ratio between obscured and unobscured
AGN, which can be checked on the CDFN and CDFS data. In Fig.~4 (right)
it is shown the ratio between the number of sources with log$N_H>22$
and log$N_H<22$ in the CDFS and CDFN as a function of redshift. Only
sources with $f_{2-10}>5 \: 10^{-16}$ \cgs and in the inner regions of
the two fields have been considered\footnote{About 1/4 of the CDFN
sample is actually selected at fluxes above $5\:10^{-15}$ \cgs (see
Barger et al. 2002).}, to get a spectroscopic completeness of $\sim
60\%$. The combined sample contains 194 sources with measured
redshift, 85 from the CDFS and 109 from the CDFN. As in Fig.~2, the
absorption column density for the CDFS sources has been calculated by
fitting the X-ray spectra with a simple absorbed power-law, fixing the
slope to $\Gamma=1.8$ when the photon statistics is low. About 64\% of
the considered CDFS sources have absorption in excess of
$10^{22}$\nh. The absorption column density for the CDFN sources has
been taken from Fig.~18 of Barger et al. (2002), who derived the $N_H$
values from the source hardness ratios (fixing the photon index to
$\Gamma=1.8$). While the redshifts considered in the CDFS subsample
are all spectroscopic, one third of the redshifts in the CDFN
subsample are photometric. Similarly to what found in the CDFS, 72\%
of the considered CDFN sources have $N_H>10^{22}$\nh. The shaded area
in Fig.~4 (right) shows the possible range covered by the ratio
between the number of sources with column density above and below
$10^{22}$\nh under the two extreme assumptions that the unidentified
sources are either all obscured or all unobscured. Although the
incompleteness is likely to increase with redshift, it is assumed to
be $40\%$ in each redshift bin. The ratio predicted by the
Franceschini et al. (2002) model, calculated at a comparable limiting
flux is also shown as a solid line. At low redshifts the predicted
ratio highly overestimates the data, while the opposite is true at
high redshifts. It is noted that the Franceschini et al. model is a
simple approximation, since only one class of obscured sources is
considered (with $N_H \sim 2\:10^{23}$\nh). In the more refined model
by Gandhi \& Fabian (2002), where several classes of sources with
different obscuration are assumed, the discrepancy is less
significant, but still the ratio between AGN with log$N_H>22$ and
log$N_H<22$ is overestimated at $z<1$.

Fig.~4 (right) indicates that at $z<1$ the ratio between AGN with
log$N_H>22$ and log$N_H<22$ is lower than $\sim 3$, suggesting that
the low-z peak in the redshift distribution is not due exclusively to
obscured sources. Since the XLF of unobscured AGN is not properly
sampled by ROSAT at low luminosities and moderate redshifts
($10^{42}$\lum at $z\sim 1$), a regime now accessible to Chandra, the
assumed extrapolations might not be correct (preliminary results
suggest this is indeed the case; see Cowie et al. 2003 and Hasinger et
al. 2003) and a new determination of the AGN XLF is therefore needed.

\section*{CONCLUSIONS}

The deep surveys have finally established that the cosmic X-ray
background in the 2-10 keV band is produced by the integrated emission
of obscured and unobscured AGN, confirming to the first order the main
prediction of population synthesis models. Now it is possible to put
tighter constraints to the model parameter space and study in detail
the AGN evolution down to luminosities of $10^{42}$\lum at $z\sim 1$,
checking if previous assumptions were correct. The AGN redshift
distribution observed in the deep surveys seems to peak at $z<1$,
while standard synthesis models were expecting a peak at
$z=1.3-1.5$. Assuming that obscured AGN evolve very quickly up to
$z<1$ does solve the discrepancy in the redshift distribution but
predicts too many obscured AGN among low-z sources. A new
determination of the XLF of unobscured AGN in the regime not covered
by previous shallower surveys is therefore needed, which has then to
be implemented into the XRB synthesis models.

\section*{ACKNOWLEDGMENTS}
I gratefully acknowledge all the members of the Chandra Deep Field
South collaboration.

\end{document}